\documentclass[journal,transmag]{IEEEtran}
\usepackage[pdftex]{graphicx}
\graphicspath{{pdf/}{figures/}}
\usepackage{svg}
\usepackage{makecell}
\usepackage{multirow}
\usepackage[ruled,vlined,algo2e]{algorithm2e}
\usepackage{algorithm}
\usepackage{algorithmic}
\usepackage[numbers,sort&compress]{natbib}
\usepackage{xcolor,soul,framed} 
\colorlet{shadecolor}{yellow}
\usepackage{array}
\usepackage{mdwmath}
\usepackage{mdwtab}
\usepackage{threeparttable}
\usepackage{eqparbox}
\usepackage{url}
\usepackage{svg}
\usepackage{balance}
\usepackage[cmex10]{amsmath}
\usepackage [colorlinks,linkcolor=blue] {hyperref}
\usepackage[justification=centering]{caption}
\usepackage{url}
\ifCLASSINFOpdf
\else
\fi

\usepackage{hyperref}
\hypersetup{colorlinks=true, linkcolor=blue, anchorcolor=blue, citecolor=blue, filecolor=blue, urlcolor=blue}

\begin{document}

 \title{DPCIPI: A pre-trained deep learning model for predicting cross-immunity between drifted strains of Influenza A/H3N2}
 \author{\IEEEauthorblockN{Yiming Du,
 Zhuotian Li,
 Qian He,
 Thomas Wetere Tulu,
 Kei Hang Katie Chan,
 Lin Wang,
 Sen Pei,\\
 Zhanwei Du,
 Xiao-Ke Xu$^{*}$ and
 Xiao Fan Liu$^{*}$}
 \thanks{Manuscript received xxx; revised xxx. Corresponding authors: Xiao-Ke Xu and Xiao Fan Liu.}
\IEEEcompsocitemizethanks{\IEEEcompsocthanksitem Yiming Du is with the Department of System Engineering and Engineering Management, The Chinese University of Hong Kong, Hong Kong SAR, China (e-mail: ydu@se.cuhk.edu.hk).}
\IEEEcompsocitemizethanks{\IEEEcompsocthanksitem Xiao-Ke Xu is with the Computational Communication Research Center, Beijing Normal University, Zhuhai 519087, China and the School of Journalism and Communication, Beijing Normal University, Beijing 100875, China (e-mail: xuxiaoke@foxmail.com).}
\IEEEcompsocitemizethanks{\IEEEcompsocthanksitem Xiao Fan Liu is with the Web Mining Laboratory, Department of Media and Communication, City University of Hong Kong, Hong Kong SAR, China (e-mail: xf.liu@cityu.edu.hk).} 
\IEEEcompsocitemizethanks{\IEEEcompsocthanksitem Zhuotian Li is with the Division of Medical Science, Faculty of Medicine, The Chinese University of Hong Kong, Hong Kong SAR, China.} 
\IEEEcompsocitemizethanks{\IEEEcompsocthanksitem Qian He, Thomas Wetere Tulu and Kei Hang Katie are with the Department of Biomedical Sciences, City University of Hong Kong, Hong Kong, China}
\IEEEcompsocitemizethanks{\IEEEcompsocthanksitem Lin Wang is with Department of Genetics, University of Cambridge, London, UK.} 
\IEEEcompsocitemizethanks{\IEEEcompsocthanksitem Sen Pei is with Department of Environmental Health Sciences at Mailman School of Public Health, Columbia University, New York, US}
\IEEEcompsocitemizethanks{\IEEEcompsocthanksitem Zhanwei Du is with Division of Epidemiology and Biostatistics, LKS Faculty of Medicine, The University of Hong Kong, Hong Kong SAR, China}}

\markboth{IEEE Journal of Biomedical and Health Informatics}{Du \MakeLowercase{\textit{et al.}}: DPCIPI for estimation of cross-immunity between drifted strains}

\maketitle
\begin{abstract}
Predicting cross-immunity between viral strains is vital for public health surveillance and vaccine development. Traditional neural network methods, such as BiLSTM, could be ineffective due to the lack of lab data for model training and the overshadowing of crucial features within sequence concatenation. The current work proposes a less data-consuming model incorporating a pre-trained gene sequence model and a mutual information inference operator. Our methodology utilizes gene alignment and deduplication algorithms to preprocess gene sequences, enhancing the model's capacity to discern and focus on distinctions among input gene pairs. The model, i.e., DNA Pretrained Cross-Immunity Protection Inference model (DPCIPI), outperforms state-of-the-art (SOTA) models in predicting hemagglutination inhibition titer from influenza viral gene sequences only. Improvement in binary cross-immunity prediction is 1.58\% in F1, 2.34\% in precision, 1.57\% in recall, and 1.57\% in Accuracy. For multilevel cross-immunity improvements, the improvement is 2.12\% in F1, 3.50\% in precision, 2.19\% in recall, and 2.19\% in Accuracy. Our study highlights the potential of pre-trained gene models in revolutionizing gene sequence-related prediction tasks. With more gene sequence data being harnessed and larger models trained, we foresee a significant impact of pre-trained models on clinical and public health applications.
\end{abstract}
\begin{IEEEkeywords}
Cross-immunity prediction, pre-trained model, deep learning, influenza strains, hemagglutination inhibition.
\end{IEEEkeywords}

\maketitle

\IEEEdisplaynontitleabstractindextext

\IEEEpeerreviewmaketitle

\section{Introduction}
\IEEEPARstart{S}{easonal} influenza infects up to a billion people worldwide yearly, causing millions of severe cases and up to 650,000 deaths \cite{bullard_epitope-optimized_2022}. Vaccination before the onset of a seasonal epidemic is the most effective way of prevention and control \cite{davenport_current_1962, earn_ecology_2002}. Unfortunately, influenza viruses can evade immune system control through a process known as antigenic drift \cite{finlay_anti-immunology_2006, vossen_viral_2002}. The efficacy of previous influenza vaccines to protect against drifting strains depends on the antigenic similarity between the vaccine and epidemic strains \cite{thompson_increased_2019}. Hemagglutinin (HA), a surface glycoprotein of the influenza virus, is primarily responsible for triggering the immune response, specifically its subunit, HA1. The measurement of Hemagglutination Inhibition (HI) titer, obtained from the assay, evaluates the degree of antigenic similarity \cite{kirchenbaum_competitive_2021, potter_determinants_1979}. The traditional HI experiment involves preparing antibodies, diluting the antibodies, reacting with antigens, and finally being tested by red blood cells \cite{sawant2023h3n2}. Due to the pairwise reaction of antigens and antibodies, the workload is enormous in the face of thousands of mutated epidemic strains every year and it is a labor-intensive and time-consuming process. What's worse, taking the H3N2 influenza virus as an example, thousands of strains are submitted annually. Millions of HI tests would be required to thoroughly assess antigenic variations. Therefore, it is imperative to explore computational models that can assess antigenic variation without relying solely on conducting HI tests.

Studies have found that point mutations or deletions in the HA1 gene sequence can lead to immune failure, reducing the efficacy of the vaccine \cite{asaduzzaman_estimation_2018, fulvini_ha1_2021, smith_mapping_2004,nyangau_genetic_2020}. Intuitively, one can utilize neural networks to predict cross-immunity using HA1 gene sequences, which have demonstrated their capabilities in the past decade \cite{alharbi2022review, al2019cnn, zhang2017titer}. For example, LSTM-based models demonstrated the capability to forecast mutation probabilities \cite{lim2020evolstm}, predict DNA-protein binding \cite{zhang2020deepsite}, and produce new reasonable molecules \cite{wang2021improving}. CNN-based models \cite{kim2014convolutional} exhibited proficiency in variant detection \cite{poplin2018universal}, cancer type prediction \cite{mostavi2020convolutional}, and gene expression dynamic profiles classification \cite{mitra2021rvagene}. BiLSTM-based models \cite{huang2015bidirectional} showcased the ability to predict viral escape \cite{hie_learning_2021}, cancer types \cite{metipatil2023efficient}, and genetic disorders \cite{nandhini2023optimal}.

Despite no prediction model that utilizes neural networks to predict HI titers, simply applying existing models may also have limitations \cite{Xu2023SupplementaryFO}. First, the performance of these models is highly dependent on training data \cite{alharbi2022review}. The time-consuming nature of the HI assay \cite{spackman2020hemagglutination} restricts the amount of available data for training the neural network. Consequently, this insufficiency hinders the models' ability to learn hidden patterns in the gene. Second, current models do not consider mutual information between genes \cite{al2019cnn}. Typically, the two gene sequences input into the model are concatenated, treating them as a single sequence during training. However, this serialized input order hampers the model's ability to capture differences among genes and disregards the unique characteristics of each individual gene.

\begin{figure}[t]

\includegraphics[width=0.48\textwidth]{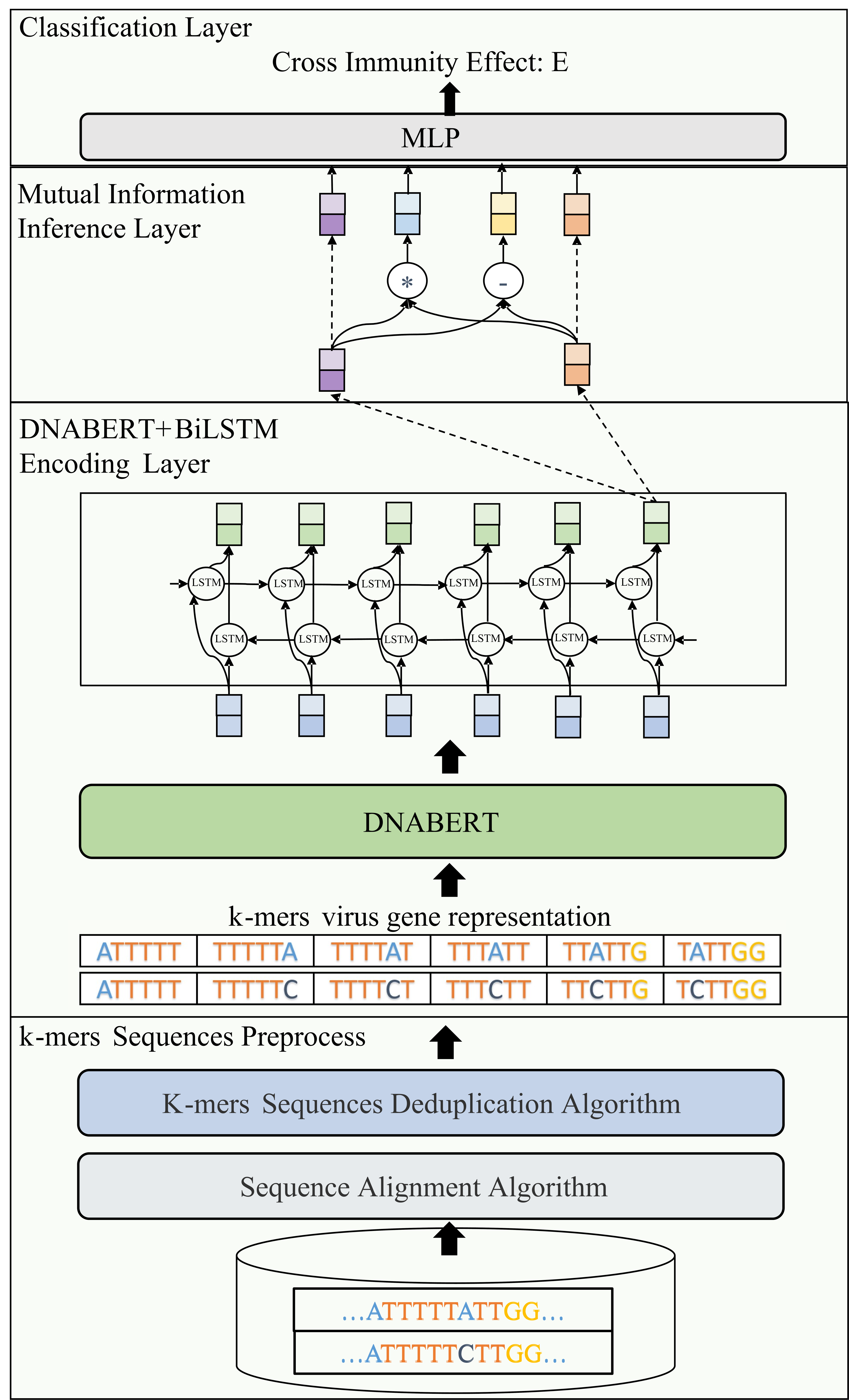}
\captionsetup{justification=justified}
\caption{The DNA Pretrained Cross-immunity Protection Inference-Model (DPCIPI) framework.}
\label{fig1:framework}
\end{figure}

This research bridges the gaps by proposing a novel model, i.e., the DNA Pretrained Cross Immunity Protection Inference Model (DPCIPI). As illustrated in Fig. \ref{fig1:framework}, the DPCIPI model consists of four key procedures, including (1) the $k$-mer sequence preprocess which converts influenza gene sequences into $k$-mers ($k$-mers are substrings of length k contained within a gene sequence), (2) a pre-trained model-based encoding layer for finding the embeddings of the $k$-mers, (3) a mutual information inference layer for capturing the between the reference and test viruses, and (4) a classification layer for inferring the cross-immunity labels.

Our contributions are in the design of the pre-trained model encoding layer and mutual information inference layer. Firstly, adaptive DNABERT initialization in the pre-trained model encoding layer is proposed to improve data insufficiency by leveraging a pre-trained model on human genes. We preprocess gene sequences into the $k$-mer format adaptable to DNABERT, and DPCIPI initializes the $k$-mer vectors with weights from DNABERT. In cases where certain uncommon $k$-mers are absent in DNABERT's weights, we represent them by averaging the surrounding $k$-mer vectors, skillfully avoiding random initialization. Secondly, we adopt a mutual information inference operator, which has proven effective in natural language inference tasks \cite{wang_bilateral_2010}. The operation operator could reveal significant dissimilarities and resemblances between sequences. By applying arithmetic operations within the mutual information inference operator to two input gene vectors, we efficiently capture feature information, distinguish genetic variations, and identify distinct traits within gene sequences, eliminating missing mutual information. Third, compared to contemporary state-of-the-art models, DPCIPI exhibits a superior precise prediction of the analogous expression between the reference and test viruses.

The remainder of this study is structured as follows: Section \ref{related work} reviews computer parameter models, pre-trained models, and co-expression methods for gene sequence analysis. Section \ref{preliminaries} lays out the problem definition, data construction, and metrics. Section \ref{method} elaborates on the proposed DPCIPI framework. Section \ref{experiments} presents experimental evaluations. Section \ref{conclusion and future work} concludes and outlines future research directions.

\section{Related Work}\label{related work}

\subsection{Computer Parameter Models}

\subsubsection{Statistical Learning Methods}
Statistical learning methods are essential in analyzing genetic sequences, particularly in elucidating the resemblances between genes to discern their interconnected functionalities. Hooper et al. \cite{hooper_prediction_2000} introduced a two-stage logistic regression approach to predict genetic structures within eukaryotic DNA. Yang et al. \cite{yang_novel_2011} employed a $k$-mer mixture logistic regression model, delineating the susceptibility of DNA methylation across diverse cell types. Nevertheless, the intricate regulatory interplays among genes often exhibit nonlinear or nonmonotonic characteristics, thereby posing a challenge for their explication through linear models.  To surmount this challenge, the perceptron models have been advocated to capture the intricate regulatory relationships inherent in genes derived from single-cell RNA-seq data \cite{luo2022signet}. In a different vein, decision trees have also been harnessed to prognosticate the multifaceted functionalities of open reading frames \cite{schietgat2010predicting}. However, it is noteworthy that these methodologies tend to impose constrictive assumptions concerning gene expression dynamics, thereby limiting their capacity to effectively capture complex gene interactions.

\subsubsection{Neural Network Methods}
Various neural network approaches have been proposed to address the intricacies of gene sequence analysis, encompassing recurrent neural networks (RNNs) \cite{zaremba2014recurrent}, Long Short-Term Memory (LSTM) networks \cite{hochreiter1997long}, Gated Recurrent Unit (GRU) networks \cite{chung2014empirical}, convolutional neural networks (CNNs) \cite{kim2014convolutional}, and Bidirectional LSTMs (BiLSTMs) \cite{huang2015bidirectional}. Notably, the Recurrent Neural Network-based Gene Regulatory Network (RNN-GRN) \cite{raza2016recurrent} integrates the generalized extended Kalman model to introduce non-linear features for gene analysis. A recent approach EvoLSTM \cite{lim2020evolstm} leverages LSTM architecture to simulate gene sequence evolution, capturing intricate mutational context dependencies within genes. In gene variation analysis, the DeepVariant model \cite{poplin2018universal}, a deep CNN-based architecture, has demonstrated proficiency in calling genetic variations within aligned sequencing gene data. This model exhibits generalizability across diverse genome builds and mammalian species \cite{poplin2018universal}. This work draws an analogy between mutations and word alterations in a sentence, both preserving grammaticality while altering the meaning. To address these mutations, a BiLSTM-based model \cite{hie_learning_2021} has been proposed to identify escape mutations. Nonetheless, it is imperative to acknowledge the inherent limitations of these methodologies, primarily due to their sensitivity to the scale of accessible data. The intricate nature of certain gene patterns continues to present challenges for comprehensive assimilation within the current framework of these approaches.

\subsection{Pre-trained Models}
Pre-trained models such as BERT \cite{devlin_bert_2018}, GPT-2 \cite{radford_language_2019}, and RoBERTa \cite{liu_roberta_2019} have made significant strides in the field of natural language processing, showcasing their proficiency in capturing intricate patterns from training corpora and aptly generalizing to specific tasks. These strengths have also been recognized and harnessed in genetics to create specialized models. DNABERT \cite{ji_dnabert_2021}, originally designed for pre-training on human DNA sequences, shows impressive adaptability, allowing it to be used for various tasks such as gene prediction, variant calling, and sequence alignment. Particularly noteworthy is DNABERT's capacity for seamless adaptation to diverse genomes. This adaptability is highlighted in its outperformance, as observed in a comprehensive evaluation involving 78 mouse ENCODE ChIP-seq datasets. Remarkably, even when pre-trained on the human genome \cite{stamatoyannopoulos2012encyclopedia}, DNABERT surpasses the efficacy of CNN, CNN + LSTM, CNN + GRU, and randomly initialized DNABERT. This robust and superior performance underscores its inherent cross-domain prowess. Expanding its utility further, DNABERT demonstrates competence in handling cross-linking and immunoprecipitation (CLIP-seq) data, thereby facilitating predictions pertaining to RNA-binding protein (RBP) binding preferences \cite{gerstberger2014census}. This wide-ranging applicability extends its potential usage within the viral domain.

\subsection{Co-expression Methods}

Co-expression methods represent widely employed techniques in analyzing gene expression data, classifiable into two principal categories: correlation coefficients and mutual information measures. Among correlation coefficients, approaches such as Weighted Gene Co-expression Network Analysis (WGCNA) \cite{langfelder2008wgcna} stand out for their ability to discern potential biomarkers or therapeutic targets. Pearson correlation \cite{cohen2009pearson}, for instance, is leveraged to distill gene features from microarray gene expression data characterized by high dimensionality and limited samples \cite{9712115}. However, correlation coefficient methods encounter challenges rooted in multicollinearity, particularly when variables exhibit pronounced interdependence. This can complicate the disentanglement of individual contributions within gene analysis. Conversely, mutual information measures like Algorithm for the Reconstruction of Accurate Cellular Networks (ARACNE) \cite{barman2017novel} possess the capability to capture non-linear gene expressions. Yet, they grapple with issues encompassing discretization, sample size, and computational intensity during gene analysis. Notably, conventional co-expression methodologies encounter impediments when confronted with high-dimensional vector spaces. An effective strategy to infer mutual information, drawn from the realm of natural language processing, pertains to utilizing Natural Language Inference (NLI) technology \cite{chen2016enhanced, mou_natural_2015}. This technology can be applied to delve into the mutual information inherent in gene representation vectors engendered by neural network models. By employing operations such as multiplication, subtraction, and preservation on sentence-level vectors, this approach captures intricate non-linear gene expressions. Given the inherent analogies shared between gene sequences and textual constructs, the transference of this methodology for the analysis of influenza genes is poised to yield insights of significance.

\section{Preliminaries}\label{preliminaries}

\subsection{Problem Definition}\label{problem definition}
The objective of this paper is to approach cross-immunity between viruses as a machine learning problem, specifically focusing on the cross-immunity prediction of their functional gene sequences measured by HI titers. The model, denoted as $M$, predicts the protective effect ($E$) of antibodies derived from reference viruses ($S_R$) on test viruses ($S_T$). The binary prediction problem distinguishes $E$ as present or absent (0 or 1), while the multilevel prediction problem divides $E$ into four levels (0, 1, 2, or 3). The study treats cross-immunity prediction as a classification problem and seeks to estimate the probability ($P$) of the protective effect ($E$) given the gene sequences ($S_R$ and $S_T$): 
\begin{equation}
\label{eq2}
    P (E|S_R, S_T) = M (S_R, S_T),
\end{equation}
where the machine learning model denoted as $M$. DPCIPI model is introduced as a specific instance within the broader set of models encompassed by $M$, emphasizing its significance as the primary contribution of the research.

\subsection{Dataset Construction}\label{Dataset Construction}
In this study, we compile a revised dataset sourced from Smith et al. \cite{smith_mapping_2004}, subsequently redesignated as the Virus Hemagglutination Inhibition Dataset (VHID). This compilation encompasses a total of 2,472 hemagglutination inhibition (HI) titer outcomes derived from an assemblage of 240 reference viruses and 43 test viruses. We retrieved the gene sequences of the viruses from the GenBank of the NCBI database using Accession Numbers. The number of viruses in our dataset is slightly lower than that reported in \cite{smith_mapping_2004} due to duplicated accession numbers and missing gene reports. Out of 10,320 ($240\times43$) possible reference-test virus combinations, we identified 2,472 valid ones with HI titer values, leaving 7,848 samples without HI titer values. 

To categorize the samples into positive and negative groups, we adopt a threshold of 40 \cite{kaufmann2017optimized} for the HI titer, as it is widely recognized as corresponding to a 50\% reduction in the risk of influenza \cite{black_hemagglutination_2011}. Consequently, the 2,472 valid samples are divided into 1,733 positive and 739 negative samples. In the context of the multi-level cross-immunity prediction task, we partitioned the 2,472 valid samples into four intervals using the distribution of their HI titer values. The illustration reveals that the HI titer data is divided among four ranges: [0, 40), [40, 100), [100, 1000), and [1000, 10240], encompassing 693, 372, 839, and 568 data examples, respectively. Each interval is assigned a label (0, 1, 2, or 3) indicating the level of cross-immunity. Instead of using the binary classification approach, we employed the cross-entropy loss function in this experiment.  For access to the VHID dataset, please visit \url{https://github.com/Elvin-Yiming-Du/DPCIPI_cross-immunity_prediction/tree/main/VHID}.

We consider all strains reported before 1995 as the training set and those after 1995 as the test set. This ensures that the model accounts for cross-immunity between viruses that have not emerged in the future and historical strains, rather than focusing solely on cross-protection between historical strains.

\section{Method}\label{method}

\subsection{The DPCIPI Model}
\subsubsection{$k$-mers sequence preprocess}\label{k-mers preprocess}
\begin{figure}[t]
\centering
\includegraphics[width=0.48\textwidth]{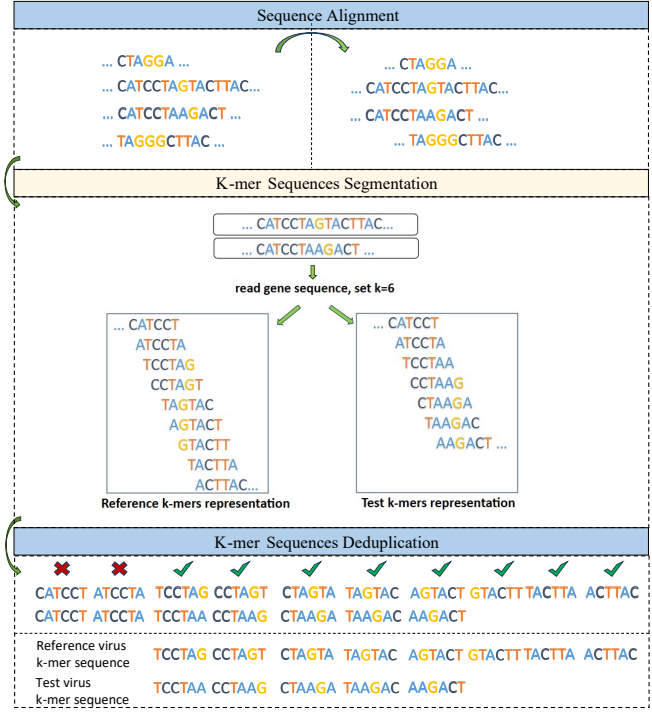}
\captionsetup{justification=justified}
\caption{A visual guide to uncovering preprocessed $k$-mer representations through gene sequence alignment and deduplication algorithm.}
\label{fig3:algorithm}
\end{figure}

To obtain the gene sequence representation inputted into the encoding layer, a preprocessing procedure is conducted on the gene sequences present within VHID, as illustrated in Fig. \ref{fig3:algorithm}.

\textbf{Gene sequence alignment}. Due to the strong similarities among influenza virus gene sequences, we begin by aligning all functional gene sequences in the VHID dataset with a common template. This alignment process helps us to pinpoint the differences between the sequences. We use the Sequence Alignment Algorithm (Algorithm \ref{algorithm_1}) to analyze a set of gene sequences denoted as $S$. The primary goal of this algorithm is to find the starting alignment positions ($D$), which indicate how each gene sequence aligns relative to the leftmost endpoint.

\begin{algorithm}[t]
{
\caption{Sequence Alignment Algorithm}
\label{algorithm_1}
\KwIn{$S$ (The unique gene sequences dict: key is virus names, value is sequences.)}
\KwOut{$D$ (The start alignment position dict: key is virus names, value is the distance between the farthest starting position and the current sequence starting position.)}
\SetKwFunction{AlignSequences}{\textbf{AlignSequences}}
\SetKwFunction{FindStartPosition}{\textbf{FindStartPosition}}
\SetKwFunction{CalculateCommonPositions}{\textbf{CalculateCommonPositions}}
\SetKwProg{Fn}{Function}{:}{}
\Fn{\AlignSequences{S}}{
$\hspace{0.5em}$Find the longest gene sequence $s_{max}$ and the corresponding length $l$ from $S$;\\
Initialize $D \gets \left[\right]$;\\
\For{each $s \in S$}{
$\enspace d \gets 0$;\\
$d \gets$ \FindStartPosition{$s.value, s_{max}.value$};
$D\left[s.key\right] \gets d$;
}
$\hspace{0.5em}$\KwRet{$D$};
}
\Fn{\FindStartPosition{$s.value, s_{max}.value$}}{
$\hspace{0.5em}$$p \gets$ the length difference between $s_{max}.value$ and $s.value$;\\
$d \gets 0$;\\
$q \gets 0$;\\
\For{$i \gets 0$ \KwTo $p$}{
{$c\_length \gets$ the common sites length between $s.value$              
and $s_{max}.value\left[i:\right];$}\\
\If{$q \leq$ $c\_length$}{
$\hspace{0.5em}$$q \gets$ the common sites length between  $s.value$ and $s_{max}.value\left[i:\right];$\\
$d \gets i;$
}
}
$\hspace{0.5em}$\KwRet{$d$};
}
}
\end{algorithm}

To achieve this, the algorithm first identifies a reference sequence, $s_{max}$, which is the most extensive typical homologous sequence. It then employs the \textbf{\texttt{FindStartPosition}} function to calculate the start alignment positions dictionary ($D$) by identifying common sites among the sequences.

\textbf{$k$-mer segmentation}. Subsequently, we extract the reference virus gene sequence $S_R$, and the test virus gene sequence $S_T$, from VHID. These sequences are then converted into representations known as $k$-mers as depicted in Figure \ref{fig3:algorithm}. $k$-mers are DNA segments consisting of consecutive nucleotides, each segment having a length of $k$. In our practical implementation, we set $k$ to 6, a choice validated for its high performance by \cite{ji_dnabert_2021}. The resulting $k$-mer sequences for the reference and test viruses are labeled as ${S_R}^{*}$ and ${S_T}^{*}$, respectively, where '${*}$' signifies the k-mer format used in this paper.

\textbf{$k$-mer sequences deduplication}. To eliminate the influence of identical locus $k$-mers on the prediction of cross-immunity, we design a $k$-mer Sequences Deduplication Algorithm (Algorithm \ref{Algorithm 2}) to remove duplicate $k$-mer from reference virus ${S_R}^{*}$ and test virus ${S_T}^{*}$ at the same locus. The algorithm consists of two main functions:
\begin{itemize}
    \item \textbf{\texttt{DeduplicationPairSequences}} aligns and fills the pair sequences while identifying and recording common positions between them. It then removes the common $k$-mers from both sequences, resulting in modified sequences ($S_{R_d}^{*}$ and $S_{T_d}^{*}$) without duplicate $k$-mers, where subscript ${d}$ indicates the $k$-mers after deduplicatoin.
    \item \textbf{\texttt{AlignmentAndFillingPairs}} aligns and fills the sequences with placeholders, based on the start alignment positions in $D$. Finally, the algorithm outputs the pre-processed reference and test virus $k$-mer sequences.
\end{itemize}

\begin{algorithm}[t]
\caption{$k$-mer Sequences Deduplication Algorithm}
\label{Algorithm 2}
\KwIn{$S_R^{*}$ (Reference virus $k$-mer sequences), $S_T^{*}$ (Test virus $k$-mer sequences), $R_{name}$ (Virus name of $S_R^{*}$), $T_{name}$ (Virus name of $S_T^{*}$), $D$ (Start alignment position dictionary),  $k$ (Size of $k$-mer)}
\KwOut{$S_{R_d}^{*}$ (Modified $S_R^{*}$ without common $k$-mers), $S_{T_d}^{*}$ (Modified $S_T^{*}$ without common $k$-mers)}

\SetKwFunction{DeduplicationPairSequences}{\textbf{DeduplicationPairSequences}}
\SetKwFunction{AlignmentAndFillingPairs}{\textbf{AlignmentAndFillingPairs}}

\SetKwProg{Fn}{Function}{:}{}

\Fn{\DeduplicationPairSequences{}}{
$\hspace{0.5em} m,n \leftarrow$ \AlignmentAndFillingPairs{$S_R^{*}, S_T^{*},\\
R_{name}, T_{name}, D, k$};\\
$l \leftarrow$ Min (the length of m, the length of n);\\
$O \leftarrow\left[\right]$, record the common position between m and n;\\
$S_{R_d}^{*}$ $\leftarrow \left[\right]$, $S_{T_d}^{*}$\textsuperscript{}$\leftarrow\left[\right]$;\\
\For{$i \leftarrow 0$ \KwTo $l$}{
\If{$m[i]==n[i]$}{
add $i$ to O;
}
}
$\hspace{0.5em}S_{R_d}^{*} \leftarrow$ delete m[o] or $k\#$ from m where o in O;\\
$S_{T_d}^{*} \leftarrow$ delete n[o] or $k\#$ from n where o in O;\\
$\hspace{0.5em}$\KwRet{$S_{R_d}^{*}$, $S_{T_d}^{*}$};
}
\Fn{\AlignmentAndFillingPairs{$S_R^{*}, S_T^{*},\\
R_{name}, T_{name}, D, k$}}{
$\hspace{0.5em}m,n \leftarrow\left[\right]$;\\
\For{$i \leftarrow 0$ \KwTo $D[s_{name}]$}{
add $k*\#$ to m;
}
$\hspace{0.5em}$\For{$j \leftarrow 0$ \KwTo $D[t_{name}]$}{
add $k*\#$ to n;
}
\enspace $m \leftarrow$ add split($S_R^{*}$) to m;\\
$n \leftarrow$ add split($S_T^{*}$) to n;\\
$\hspace{0.5em}$\KwRet{m, n};
}
\end{algorithm}

\subsubsection{The DNABERT+BiLSTM encoding layer}\label{encoding_layer}
\textbf{DNABERT} is a pre-trained gene sequence model based on BERT. It takes $k$-mers as input and obtains hidden embeddings of individual $k$-mers by unsupervised learning. The model achieves excellent performance when validated by downstream tasks, such as predicting promoter regions, which also side-by-side validates that the hidden layer vector of $k$-mers contains rich expression information of gene sequences.

Enter ${S_{R_d}}^{*}$ and ${S_{T_d}}^{*}$ , respectively, into the DNABERT pre-trained language model to obtain the initialization sequence representation:
\begin{equation}
 {X_{R_d}}^{*}=\textrm{DNABERT} ({S_{R_d}}^{*}),
\end{equation}
\begin{equation}
 {X_{T_d}}^{*}=\textrm{DNABERT} ({S_{T_d}}^{*}),
\end{equation}
where the vector ${X_{R_d}}^{*}$ = ($x_{R_d}^{*1}, x_{R_d}^{*2}, x_{R_d}^{*3}, \ldots, x_{R_d}^{*m}$) represents the $k$-mer vectors for reference virus $k$-mer sequence ${S_{R_d}}^{*}$. The value of $m$ corresponds to the length of the sequence. Similarly, the vector ${X_{T_d}}^{*}$ = ($x_{T_d}^{*1}, x_{T_d}^{*2}, x_{T_d}^{*3}, \ldots, x_{T_d}^{*n}$), representing the $k$-mer vectors for test virus $k$-mer sequence ${S_{T_d}}^{*}$. The value of $n$ corresponds to the length of this sequence. The vectors $x_{R_d}^{*i}$ and $x_{T_d}^{*j}$ serve as vector representations for the respective $k$-mers within their sequences. It is worth noting that for $k$-mers that do not appear in DNABERT, we adopt the mean value of the $k$-mer vector at the adjacent position (with a window size of 2) to initialize the $k$-mer. This approach ensures that even for unknown $k$-mers, we can still provide a suitable initialization based on neighboring $k$-mer information obtained from DNABERT. 

\textbf{BiLSTM} is a bi-directional long short-term memory neural network that can capture the sequence meaning after the recurrent encoding. The embedding vectors ${X_{R_d}}^{*}$ and ${X_{T_d}}^{*}$ are then encoded with BiLSTM into corresponding sequential embeddings $p$ and $r$:
\begin{equation}
p = \textrm{BiLSTM} ({X_{R_d}}^{*}),
\end{equation}
\begin{equation}
r = \textrm{BiLSTM} ({X_{T_d}}^{*}). 
\end{equation}

\subsubsection{The mutual information inference layer}
A \textbf{mutual information inference operator} is used to fuse the information of the two sequence-level embeddings using techniques from natural language inference \cite{mou_natural_2015}. In natural language inference, the task is to predict whether a hypothesis can be inferred by a premise, essentially using the inference results labeled in advance to analyze some similarity between two sentences. Similarly, we can use annotated HI titer markers to analyze whether the antibody generated by the reference virus to stimulate the immune system could cross-protect another test virus.

The mutual information inference operator extracts similar information between sequences by performing dot production and subtraction on vectors \cite{turney_domain_2012}. The splicing vector q is obtained from the mutual information inference operator:
\begin{equation}
\label{eq9}
q=[p; p*r; p-r; r],
\end{equation}
where $p*r$ represents the multiplication operation, $p-r$ represents the subtraction operation in the mutual information inference operator, which have a strong enhancement effect on the salient features and difference features in the $k$-mer representation vector, and the original hidden mode information of the two original $k$-mers vectors $p$ and $r$ are preserved. The final interaction information will be fed into the full neural network model to predict cross-protection.

\subsubsection{The classification layer}
In the hemagglutination inhibition test \cite{kaufmann2017optimized}, the highest dilution of hemagglutinin working fluid at which red blood cells are not completely agglutinated is used as the endpoint of determination. Since the working solution was used in twofold dilution, the experimental results showed discrete characteristics, so we used classification machine learning methods to make inferences about the similarity of cross-immunity. Put the splicing vector $q$ obtained by the mixing layer into a multi-layer perceptron neural network (MLP) to get the classification result $y'$ :
\begin{equation}
\label{eq9}
P(y'| q)=\textrm{MLP} (q). 
\end{equation}

\subsection{$k$-mer Embedding Initialized with DNABERT}

When working with gene sequences, there can be various $k$-mers that were not encountered during the pre-training of DNABERT. These unknown $k$-mers might correspond to specific genetic variations or rare sequences not included in the original training data. If these unknown $k$-mers are not appropriately initialized, it can lead to several problems. Firstly, it may result in incorrect representations, where the model assigns random or arbitrary embeddings to these $k$-mers, leading to inaccurate representations and difficulty in proper interpretation. Secondly, there can be a loss of valuable genetic information encoded in the unknown $k$-mers, hindering the model's ability to capture essential patterns and relationships in the gene sequences. Additionally, improper initialization may introduce bias in the model predictions since it lacks sufficient information about these $k$-mers, potentially leading to predictions that do not reflect the true characteristics of the gene sequences. 

To mitigate these problems, DNABERT employs the mean value of the $k$-mer vector at the adjacent position for initializing unknown $k$-mers, providing a reasonable approximation based on the context of neighboring $k$-mers. This strategy enables the model to masterfully handle previously unseen $k$-mers, preserving the integrity of the gene sequence information. Consequently, proper initialization ensures that the model can generalize expertly and make more accurate predictions on a wide range of gene sequences, including those containing new or rare $k$-mers. In summary, the initialization of $k$-mers embedding with DNABERT is crucial in capturing hidden patterns and reducing predictive performance drops due to sparse data.

\subsection{Mutual Information Inference Operators}
Drawing inspiration from natural language inference, the mutual information inference operator fuses information from two sequences by examining their similarity. Similar to how natural language inference predicts the inferability of a hypothesis from a premise, the mutual information inference operator can assess whether the antibody produced by a reference virus can cross-protect another test virus using the annotated HI titer.

The operator employs vector dot production and subtraction to extract similar information between sequences. The resulting splicing vector, denoted as $q$, is obtained through a combination of operations: multiplication ($p*r$) and subtraction ($p-r$). These operations enhance the salient features and differences in the $k$-mer representation vector while preserving the original hidden-mode information of the two original $k$-mer vectors ($p$ and $r$). The interaction information derived from this process is then fed into a full neural network model to predict cross-immunity.

\section{Experiments}\label{experiments}

\subsection{Baseline Models}

\subsubsection{Statistical Methods}
The statistical learning methods mainly use the external representation features of $k$-mers gene sequences to make cross-protective predictions directly. 

\textbf{Logistic Regression (LR)} is often used for classification when the target is categorical. In this paper, logistic regression uses the similarity scores between reference virus and test virus to predict cross-immunity, the similarity score is calculated as: 
\begin{equation}
 Sim ({S_R}, {S_T})= d({S_R},{S_T})/((L_R+L_T)/2),
\end{equation}
where $Sim ({S_R}, {S_T})$ is the similarity between reference virus ${S_R}$ and test virus ${S_T}$. $L_R$ and $L_T$ are the lengths of $S_R$ and $S_T$, respectively. The calculation of $d(S_R, S_T)$ must be aligned according to the Sequence Alignment Algorithm (as shown in Algorithm \ref{algorithm_1}) before the calculation. Moreover, the prediction function is as:
\begin{equation}
y'=\frac{1}{(1+exp (-w*Sim(S_R, S_T)-b))},
\end{equation}
where $y'$ is the predicted result coming from LR.

\textbf{Perceptron} is a supervised and classical binary classification model. In this case, the similarity of the gene sequence is used as a feature to predict cross-immunity. The input of perceptron is the linear combination of the weighted similarity between the reference virus and test virus and biases, and the sign function is used to predict cross-immunity:
\begin{equation}
 y'=sign(w*Sim(S_R, S_T)+b).
\end{equation}

\textbf{Decision Tree (DTree)} uses the branching method to classify the similarity score, and it can be used in binary and multi-label classification.

There are various methods for calculating the similarity between gene sequences. Changing the statistical learning-based similarity calculation method to the vector similarity calculation between gene sequence embeddings (\textbf{GSE}) can improve the performance in predicting cross-immunity, where GSE represents the sequence-level representation vector of $k$-mers and is treated as a cumulative average of the vectors of all $k$-mers in the sequence as:
\begin{equation}
 {p}^{'}={\frac{1}{n}\sum_{i=1}^n{X_{R_d}}^*},
\end{equation}
\begin{equation}
 {r}^{'}={\frac{1}{m}\sum_{i=1}^m{X_{T_d}}^*}. 
\end{equation}
Then we get LR-GSE, Perceptron-GSE, and DTree-GSE models.

\begin{table*}[ht]
\setlength{\tabcolsep}{4pt}
\renewcommand{\arraystretch}{1.2}
\captionsetup{justification=justified}
\caption{Comparison of the performance across statistic learning-based models (LR, Perceptron, DTree), neural network-based models (NN, CNN, BiLSTM), and DPCIPI models. The results include binary and multi-level cross-immunity prediction performance tested on the VHID dataset in the metric of Accuracy, Weighted F1, Weighted Precision, and Weighted Recall. `Improvement' indicates the relative improvement against the best baseline performance.}
\centering
\resizebox{18cm}{!}{
\begin{tabular}{clccccccccccc}
\hline
Task & \multicolumn{1}{c}{Metric} & \multicolumn{1}{c}{LR} & \multicolumn{1}{c}{LR-GSE} & \multicolumn{1}{c}{Perceptron} & \multicolumn{1}{c}{Perceptron-GSE} & \multicolumn{1}{c}{DTree} & \multicolumn{1}{c}{DTree-GSE} & \multicolumn{1}{c}{NN} & \multicolumn{1}{c}{CNN} & \multicolumn{1}{c}{BiLSTM} & \multicolumn{1}{c}{DPCIPI} & \multicolumn{1}{c}{Improvement} \\ \hline
\multirow{4}{*}{\begin{tabular}[c]{@{}c@{}}Binary \\ cross-immunity \\ prediction\end{tabular}}     & Weighted F1 & 65.22 & 60.82  & 72.31 & 69.39 & 80.38 & 79.08 & 80.35 & 81.88 & \underline{86.56} & \textbf{88.14} & \textbf{+1.58$\%$} \\                                    & Weighted Precision & 75.71  & 71.27  & 65.39    & 65.31 & 81.45 & 78.64 & 87.13 & 81.47 & \underline{88.06} & \textbf{90.40} & \textbf{+2.34\%}\\
& Weighted Recall  & 61.11& 56.17 & 80.86  & 74.38 & 79.63 & 79.63 & 84.69 & 82.50 & \underline{88.12} & \textbf{89.69}  & \textbf{+1.57\%} \\                                    & Accuracy & 61.11 & 56.17 & 80.86 & 74.38 & 79.63    & 79.63 & 84.69 & 82.50 & \underline{88.12} & \textbf{89.69} & \textbf{+1.57\%} \\ \hline
\multicolumn{1}{l}{\multirow{4}{*}{\begin{tabular}[c]{@{}c@{}}Multi-level \\ cross-immunity \\ prediction\end{tabular}}} & Weighted F1 & 29.32 & 10.34   & - & - & 50.89 & 41.41 & 31.26 & 42.93 & \underline{62.59} & \textbf{64.71} & \textbf{+2.12\%}\\
\multicolumn{1}{l}{} & Weighted Precision & 14.32      & 18.23 & - & - & 51.75 & 51.06 & 24.85 & 50.07       & \underline{63.75} & \textbf{67.25} & \textbf{+3.50\%} \\
\multicolumn{1}{l}{}                                   & Weighted Recall& 29.32 & 18.23 & - & - & 51.54      & 43.21 & 31.26 & 45.62 & \underline{62.50} & \textbf{64.69}  & \textbf{+2.19\%}\\
\multicolumn{1}{l}{} & Accuracy & 29.32 & 18.23 & -    & - & 51.54 & 43.21 & 31.26 & 45.62 & \underline{62.50} & \textbf{64.69}   & \textbf{+2.19\%}\\ \hline
\end{tabular}
}
\label{table 1}
\end{table*}

\subsubsection{Neural Network Methods}
The neural network methods encode $k$-mers as the gene vector and predict the cross-immunity effect.

\textbf{Neural Network (NN)} \cite{vohradsky2001neural} can be regarded as a classical classification model. In this task, we use 1NN. It concatenates gene sequences and predicts the cross-immunity effect.

\textbf{Convolutional neural network (CNN)} \cite{chen_convolutional_2015} is a classical neural network model that captures the features of sentences or images and is widely used in the field of natural language processing \cite{khosla_emotionx-ar_2018} and image recognition \cite{he_deep_2016, kipf_semi-supervised_2016}. The method of encoding the $k$-mers is the same as in section \ref{encoding_layer}.

\textbf{Bi-directional Long Short-Term Memory (BiLSTM)} \cite{chen_convolutional_2015} is good at long-term dependencies in input data, which is also an important component in DPCIPI. When regarded as a baseline model, BiLSTM concatenates two gene sequences directly and predicts the cross-immunity effect.

\subsection{Metrics}\label{Metrics}
We use Accuracy, Weighted Precision, Weighted Recall, and Weighted F1 \cite{grandini2020metrics} for evaluating the classifications. Specifically,  Weighted Precision introduces a weighting mechanism inversely correlated with the number of positive cross-immunity samples. This intricate weighting scheme effectively imposes stricter penalties for false positives within the positive cross-immunity class. Meanwhile, Weighted Recall assumes a position of heightened significance by prioritizing the recall performance within the positive cross-immunity class. The Weighted F1 score, an amalgamation of precision and recall, considers the intricate fabric of class distribution. This score aims to balance precision and recall for both classes. By adopting the Weighted Precision, Weighted Recall, and Weighted F1 score measures, we effectively alleviate the implications of imbalanced HI titer examples on the overall dependability of our experimental findings.

\subsection{Implementation}
All experiments were implemented on our server with 512 G memory and 4 Nvidia 3090 graphics cards. PyTorch \cite{paszke_pytorch_2019}, PyTorch geometric \cite{fey_fast_2019}, and the DNABERT library are used to conduct the experiments. We train DPCIPI for 50 epochs under the settings batch size = 10 and learning rate = 0.0001. 

\subsection{Results}

\subsubsection{Binary Cross-immunity Prediction}

Table \ref{table 1} presents the binary cross-immunity prediction results obtained from statistical learning methods, neural network models, and DPICPI. The table shows that DPCIPI achieves 90. 40\% in the precision metric, indicating that the model has a high degree of confidence in predicting cross-immunity. The model has achieved 1.59\%, 2.34\%, 1.57\%, and 1.57\% improvements in Weighted F1, Weighted Precision, Weighted Recall, and Accuracy, respectively, to the best-performing baseline model BiLSTM. It also achieves a performance improvement of 22. 92\%, 15. 83\%, and 7. 76\% over statistical learning methods (logistic regression, perceptron, and decision tree) in the Weighted F1 metric, respectively. 

We also found that when compared to the Eulerian distance-based similarity score calculation, gene sequence embedding (GSE) achieves the worse performance in Logistic Regression, Perceptron, and Decision tree models. It indicates that the direct use of gene sequence embedding on the similarity calculation is risky. Differently, the neural network-based models, such as DPCIPI, improved tremendously in all metrics compared to the conventional methods such as logistic regression and perceptron. Besides, we also found that using a decision tree with a max depth of 5 can achieve comparable performance to NN.

\subsubsection{Multi-level Cross-immunity Classification}

Table \ref{table 1} additionally displays the results of multi-level cross-immunity prediction obtained through statistical learning methods, neural network models, and DPICPI. DPCIPI again achieves the best performance on the metric of Accuracy (64.69$\%$), Weighted F1 (64.71$\%$), Weighted Precision (67.25$\%$), and Weighted Recall (64.69$\%$) which far surpasses other models. Compared to the best-performing BiLSTM, DPCIPI achieves a 2.12\%, 3.50\%, 2.19\%, and 2.19\% improvement in the Weighted F1, Weighted Precision, Weighted Recall, and Accuracy metrics. 

\begin{table*}[]
\renewcommand{\arraystretch}{1.2}
\setlength{\tabcolsep}{4pt}
\captionsetup{justification=justified}
\caption{Comparison of the performance of DNABERT Initialization across CNN, BiLSTM and DPCIPI models. The results includes binary and multi-level cross-immunity prediction performance tested on VHID dataset in the metric of Accuracy, Weighted F1, Weighted Precision and Weighted Recall. 'Improvement' indicates the relative improvement against the model without DNABERT Initialization. '@' indicates a concatenation operation.}
\centering
\resizebox{18cm}{!}{
\begin{tabular}{clccccccccc}
\hline
Task  & \multicolumn{1}{c}{Metric} & \multicolumn{3}{c}{CNN} & \multicolumn{3}{c}{BiLSTM} & \multicolumn{3}{c}{DPCIPI}  \\ \hline
Initialization Settings  & \multicolumn{1}{c}{}  & \multicolumn{1}{c}{@/(\%)} & \multicolumn{1}{c}{@DNABERT} & \multicolumn{1}{c}{Improvement} & \multicolumn{1}{c}{@/(\%)} & \multicolumn{1}{c}{@DNABERT} & \multicolumn{1}{c}{Improvement} & \multicolumn{1}{c}{@/(\%)} & \multicolumn{1}{c}{@DNABERT} & \multicolumn{1}{c}{Improvement} \\ \hline
\multirow{4}{*}{\begin{tabular}[c]{@{}c@{}}Binary\\ cross-immunity\\ prediction\end{tabular}} &  Weighted F1& 72.41 & 81.88        & \textbf{+9.474\%} & 86.56  & 86.94     & \textbf{+0.38\%}    & 86.97  & 88.14     & \textbf{+1.17\%}    \\
&  Weighted Precision  & 65.61  & 81.47   & \textbf{+15.86\%}  & 88.06  & 88.34  & \textbf{+0.28\% }& 88.60  & 90.40& \textbf{+1.8\%}  \\
&  Weighted Recall  & 80.94  & 82.50     & \textbf{+1.56\%}    & 88.12  & 88.44     & \textbf{+0.32\%}  & 89.06  & 89.69     & \textbf{+0.63\%} \\
& Accuracy   & 80.94  & 82.50  & \textbf{+1.56\%}    & 88.12  & 88.44  & \textbf{+0.32\%}   & 89.06  & 89.69     & \textbf{+0.63\% }   \\ \hline
\multirow{4}{*}{\begin{tabular}[c]{@{}c@{}}Multi-level\\ cross-immunity\\ prediction\end{tabular}} &  Weighted F1 & 54.38  & 46.85     & \textbf{-7.53\%}    & 62.59  & 63.14     & \textbf{+0.55\% }   & 62.59  & 64.71     & \textbf{+2.12\%}    \\      &  Weighted Precision         & 54.65  & 50.07     & \textbf{-4.58\% }   & 63.75  & 65.61     & \textbf{+1.86\%}    & 63.75  & 67.25     & \textbf{+3.5\%}     \\      &  Weighted Recall            & 55.63  & 45.62     & \textbf{-10.01\% }   & 62.50  & 64.06     & +1.56\%    & 62.50  & 64.69     & +2.19\%    \\     & Accuracy                   & 55.63  & 45.62     & \textbf{-10.01\%}    & 62.50  & 64.06     & \textbf{+1.56\% }   & 62.50  & 64.69     & \textbf{+2.19\% }   \\ \hline
\end{tabular}
}
\label{table 2}
\end{table*}

\begin{figure}[t]
\centering
\includegraphics[width=0.50\textwidth]{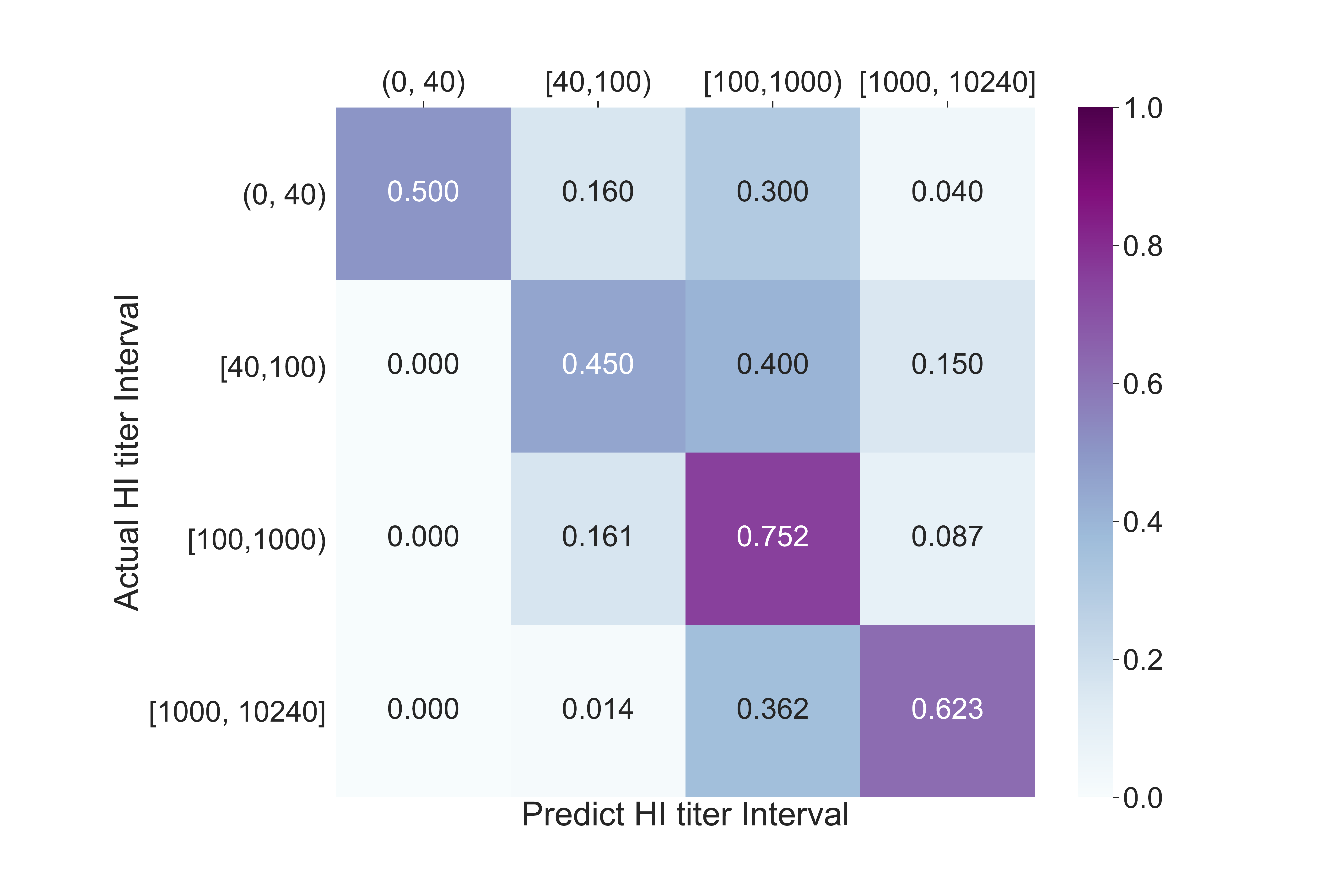}
\captionsetup{justification=justified}
\caption{The Confusion Matrix of the cross-immunity prediction on DPCIPI.}
\label{fig3:confusion_matrix}
\end{figure}

The confusion matrix in Fig. \ref{fig3:confusion_matrix} provides a straightforward depiction of the outcome. The vertical axis signifies the actual HI titer values, while the horizontal axis denotes the predicted HI titer values. The numerical entries within the heatmap cells reflect the extent of alignment in the model's predictions, ranging on a scale from 0 to 1. The outcomes showcase a notable consistency between the predicted and actual HI titer values.

Unlike LR, LR-GSE, and NN models, which have achieved poor performance (less than 30$\%$ in all metrics), the decision tree (DTree) again achieves comparable performance to the convolutional neural network model (CNN) (nearly 50$\%$). This indicates that the cut points in decision trees perform well for multi-classification tasks.

\subsubsection{DNABERT Embedding Ablation}
Pre-trained models, such as DNABERT, have shown significant progress in understanding the complex polysemy and semantic relationship between genes \cite{iuchi_representation_2021, wang_pre-trained_2021}. However, DNABERT was trained on the human genome. To evaluate the validity of DNABERT embeddings, we replaced the DNABERT embedding initialization with random embedding initialization in the model CNN, BiLSTM, and DPCIPI.

Table \ref{table 2} indicates that following initialization with DNABERT, CNN, BiLSTM, and DPCIPI have each demonstrated varying degrees of performance enhancement. In particular, the CNN model exhibits the most significant improvement, achieving an increase in precision of more than 10\% in the prediction of binary cross-immunity. On the contrary, in the multilevel cross-immunity prediction, the performance of CNN witnesses a notable decline following initialization with DNABERT but BiLSTM and DPCIPI demonstrate performance enhancements. Overall, the results demonstrate that DNABERT has captured hidden genetic patterns of virus strains and can make a significant impact on the prediction of the model.

\subsubsection{Mutual Information Inference Operator Ablation}
The incorporation of a mutual information inference (MII) operator within the model draws inspiration from the notion of a hybrid layer in natural language inference. This technique involves the fusion of two distinct word sequences to ascertain the veracity of a hypothesis ??? whether it is true (entailment), false (contradiction), or inconclusive (neutral) in relation to a given premise. To evaluate the efficacy of the MII operator within CNN, BiLSTM, and DPCIPI models, we conducted comprehensive comparative experiments. These experiments encompassed scenarios both with and without the MII operator, spanning binary cross-immunity prediction and multi-level cross-immunity prediction tasks.

\begin{table*}[]
\renewcommand{\arraystretch}{1.2}
\setlength{\tabcolsep}{6pt}
\caption{Comparison of the performance of MII operators across CNN, BiLSTM and DPCIPI models. The results include binary and multilevel cross-immunity prediction performance tested on VHID dataset in the metric of Accuracy, Weighted F1, Weighted Precision and Weighted Recall. 'Improvement' indicates the relative improvement against the model without MII operator. '@' indicates a concatenation operation}
\centering
\resizebox{18cm}{!}{
\begin{tabular}{clccccccccc}
\hline
Task & \multicolumn{1}{c}{Metric} & \multicolumn{3}{c}{CNN}                                                                 & \multicolumn{3}{c}{BiLSTM}                                                             & \multicolumn{3}{c}{DPCIPI}                                                             \\ \hline
Operator Settings &  & \multicolumn{1}{l}{@/(\%)} & \multicolumn{1}{l}{@MII(\%)} & Improvement & \multicolumn{1}{l}{@/(\%)} & \multicolumn{1}{l}{@MII(\%)} & Improvement & \multicolumn{1}{l}{@/(\%)} & \multicolumn{1}{l}{@MII(\%)} & Improvement \\ \hline
\multirow{4}{*}{\begin{tabular}[c]{@{}c@{}}Binary \\ cross-immunity \\ prediction\end{tabular}}      & Weighted-F1 & 69.54 & 81.88 & \textbf{+12.34\%} & 86.56 & 87.12 & \textbf{+0.56\%} & 86.94 & 88.14 & \textbf{+1.20\%} \\ 
& Weighted-Precision & 66.82 & 81.47 & \textbf{+14.65\%} & 88.06 & 89.26 & \textbf{+1.32\%} & 88.34 & 90.40 & \textbf{+2.06\%} \\                               & Weighted-Recall & 72.81 & 82.50 & \textbf{+9.69\%}  & 88.12 & 88.75 & \textbf{+0.63\%} & 88.44 & 89.69 & \textbf{+1.25\%} \\                               & Accuracy & 72.81 & 82.50 & \textbf{+9.69\%} & 88.12 & 88.75 & \textbf{+0.63\%} & 88.44 & 89.69  & \textbf{+1.25\%} \\ \hline
\multirow{4}{*}{\begin{tabular}[c]{@{}c@{}}Multi-level \\ cross-immunity \\ prediction\end{tabular}} & Weighted-F1 & 38.96    & 42.93 & \textbf{+3.97\%} & 61.61 & 62.59        & \textbf{+0.98\%} & 63.14 & 64.71 & \textbf{+1.57\%} \\                               & Weighted-Precision & 33.63 & 50.07 & \textbf{+16.44\%} & 65.45 & 63.75 & \textbf{+2.3\%}  & 65.51 & 67.25 & \textbf{+1.74\%} \\
& Weighted-Recall & 48.13 & 45.62 & \textbf{-2.51\%} & 60.94 & 62.50 & \textbf{+1.56\%} & 64.06 & 64.69 & \textbf{+0.63\%} \\
& Accuracy & 48.13 & 45.62 & \textbf{-2.51\%}& 60.94 & 62.50 & \textbf{+1.56\%} & 64.06 & 64.69  & \textbf{+0.63\%} \\ \hline
\end{tabular}
}
\label{table 3}
\end{table*}

Table \ref{table 3} presents the performance of the MII operator across CNN, BiLSTM, and DPCIPI for both binary and multi-level cross-immunity prediction tasks. The DPCIPI with MII achieves an improvement of 1.2\%, 2.06\%, 1.25\%, and 1.25\% in the binary classification task, and 1.57\%, 1.74\%, 0.63\%, and 0.63\% in the multi-level classification task in the metrics of Weighted F1, Weighted Precision, Weighted Recall, and Accuracy, compared to DPCIPI without MII. Furthermore, the CNN model with MII shows a significant increase (more than 10\%) in all four metrics compared to CNN without MII in the binary classification task. In the multi-level classification task, CNN with MII shows a decrease in Weighted Recall score (2.51\%) compared to CNN without MII. However, there is a significant increase in precision, resulting in a significant improvement in the Weighted F1 score.

\section{Conclusion and Future Work}\label{conclusion and future work}

The study introduces the DNA Pretrained Cross-Immunity Protection Inference (DPCIPI) model to predict cross-immunity between influenza virus strains using virus gene sequences. DPCIPI outperforms existing models in both binary and multi-level cross-immunity prediction tasks. In the binary task, DPCIPI shows significant improvements over BiLSTM and traditional statistical learning methods, such as logistic regression, perceptron, and decision tree. The 90.40\% precision indicates a high level of confidence in DPCIPI's cross-immunity predictions. In the multilevel cross-immunity classification task, DPCIPI again achieved the highest Accuracy, Weighted F1, Weighted Precision, and Weighted Recall. Confusion matrix analysis reveals consistent predictions with actual types.

Pre-trained models, specifically DNABERT, show a promising future in capturing genetic hidden patterns. Experiments carried out by replacing DNABERT embeddings with random embeddings demonstrated a decrease in performance across all models, highlighting the significant impact of DNABERT initialization on cross-immunity prediction. The mutual information inference (MII) operator is included in DPCIPI. Comparative experiments evaluate its effectiveness in CNN, BiLSTM, and DPCIPI. The results show that models with the MII operator outperform those without it in all metrics, indicating its crucial role in enhancing performance for binary and multilevel cross-immunity prediction tasks.

In conclusion, the findings of this study highlight the superiority of DPCIPI in the prediction of cross-immunity for influenza virus strains. The incorporation of pre-trained models and the mutual information inference operator proved to be an effective approach to performing prediction tasks using gene sequences. These findings carry significant implications for clinical and public health applications, offering highly efficient predictive methods for cross-immunity, as well as in silico simulation pathways for vaccine development, disease surveillance, and preparedness strategies. However, fragmented gene sequences used to train DNABERT result in incomplete gene sequence information due to the model's input length limitation. Future efforts will be made to train larger DNA length models with complete gene sequences.

\section*{Acknowledgment}
This work is part of the Grand Challenges ICODA pilot initiative, delivered by Health Data Research UK. This work has been supported by the Bill \& Melinda Gates Foundation and the Minderoo Foundation.

\balance
 \bibliographystyle{IEEEtran}
\bibliography{IEEEabrv,DPCIPI}

\end{document}